\documentclass[amsmath,aps,showpacs,floats,amssymb,preprintnumbers,floatfix,superscriptaddress,nofootinbib]{revtex4}

\usepackage{graphicx}
\usepackage{dcolumn}
\usepackage{bm}
\usepackage{hyperref} 
\usepackage[dvipsnames]{xcolor}
\usepackage[pdftex]{epsfig} 
\usepackage{amsmath}
\usepackage{feynmf}
\usepackage{feyn}

\def\square{\kern1pt\vbox{\hrule height 1.2pt\hbox{\vrule width 1.2pt\hskip 3pt
   \vbox{\vskip 6pt}\hskip 3pt\vrule width 0.6pt}\hrule height 0.6pt}\kern1pt}

\newcommand{\be}{\begin{equation}}
\newcommand{\ee}{\end{equation}}
\newcommand{\bea}{\begin{eqnarray}}
\newcommand{\eea}{\end{eqnarray}}
\newcommand{\nn}{\nonumber}

\begin{document}
\title{One loop corrected conformally coupled scalar mode equations during inflation}

\author{Sibel Boran}
\email{borans@itu.edu.tr}
\author{Emre Onur Kahya}
\email{eokahya@itu.edu.tr}
\affiliation{Department of Physics, Istanbul Technical University, Maslak 34469 Istanbul, Turkey}
\author{Sohyun Park}
\email{park@fzu.cz}
\affiliation{Korea Astronomy and Space Science Institute, Daejeon 34055, Korea}
\affiliation{CEICO, Institute of Physics of the Czech Academy of Sciences, Na Slovance 2, 18221 Prague 8 Czech Republic}

\date{\today}
\begin{abstract}
We employ a fully renormalized computation of the one loop contribution to the self-mass-squared of the conformally coupled (CC) scalar interacting with gravitons during inflation to study how inflationary produced gravitons affect the CC scalar evolution equation. The quantum corrected scalar mode functions turn out to get a secular growth effect, 
proportional to a logarithm of the scale factor at late times.  
\end{abstract}

\maketitle

\section{Introduction}\label{intro}

Quantum fluctuations of infrared gravitons and massless, minimally coupled (MMC) scalars are vastly amplified during inflation, and eventually comprise primordial tensor and scalar perturbations \cite{SMC}. This is the lowest order effect directly derives from the inflationary produced virtual quanta \cite{w01}. The interactions of this ensemble of virtual gravitons and scalars among themselves and other particles can be quantified by the one-particle-irreducible (1PI) 2-point function of a particular kind of particle. A generic procedure for studying these interactions is first to compute 
the renormalized contribution of MMC scalars or gravitons to the 1PI function of a particle in question, then to use it to quantum-correct the field equation of the particle.  

A number of computations involving either scalar-mediated effects or graviton-mediated effects during inflation have been made over the past decades:
\begin{itemize}
\item{
For MMC scalars Yukawa coupled to massless fermions, a growing mass is induced to fermions \cite{pw01, gp, mw}, but there is no large correction on scalars \cite{dw}. 
}
\item{
In MMC scalar quantum electrodynamics, assuming the scalars are charged, photons acquire a growing mass \cite{ptw, pw02, ptaw}, but scalars receive no significant effect \cite{kw05, kw06}.
}
\item{
For MMC scalars with a quartic self-interaction, scalars gain an increasing mass (though it remains perturbatively small) at one and two loop orders \cite{bow, ko}.
}
\item{
MMC scalars have no significant effect on dynamical gravitons at one loop order \cite{pw, LPPW}, but they induce a secular effect on the gravitational potentials of a point mass (specifically a secular decrease in the gravitational coupling $G$) \cite{PPW-1510}. One loop corrections from conformal fields to the gravitational potentials of a point mass were also found \cite{FRV-1403, WW-1508, FV-1601,FV-1701}.
}
\item{
For quantum gravity plus massless fermions, the fermion field strength gets a secular growth effect from inflationary gravitons \cite{mw05-06-08}.
}
\item{
For quantum gravity plus MMC scalars, inflationary gravitons induce no significant corrections on the scalar mode functions at one loop \cite{kw07, kw08}.
}
\item{
For quantum gravity plus electrodynamics, inflationary gravitons induce secular effects on the dynamical photons and alter the electric field of a point charge and the magnetic field of a point magnetic dipole \cite{LPW-1210, LPW-1211, LW, WW-1408, gmpw13, GMPW-1504, gmpw16}.
}
\end{itemize}
These secular effects are typically in the form of the logarithm of the scale factor $\ln(a)$ times the minuscule loop counting parameter $GH^2$ \cite{wein}.

The purpose of this paper is to work out leading loop corrections to the CC scalar mode functions from inflationary gravitons. 
We have already computed the fully renormalized, one loop contributions to the CC scalar self-mass-squared from gravitons \cite{bkp14, bkp17}. The result is summarized in Sec. \ref{precalc}.
We solve the quantum corrected scalar field equation at one loop order in Sec. \ref{effmodeeq}. We give our conclusions in Sec. \ref{dis}.
\section{Our Previous Calculation}\label{precalc}

We started with the bare Lagrangian of pure gravity plus the CC scalar
\bea\label{Lag}
\mathcal{L} \equiv -\frac12 \partial_{\mu} \phi \;
\partial_{\nu} \phi g^{\mu\nu} \sqrt{-g} 
-\frac{D-2}{8(D-1)} \phi^2 R \sqrt{-g} 
+ \frac1{16\pi G} \Bigl( R - (D-2) \Lambda \Bigr) \sqrt{-g}\;. 
\eea
Here $R$ is the Ricci scalar, $\Lambda \equiv (D-1) H^2$ is the cosmological constant with the Hubble constant $H$, and $G$ is Newton's constant. We worked on the open conformal coordinate submanifold of de Sitter space
\be
\hat{g}_{\mu\nu} = a^2(\eta) \eta_{\mu\nu}\;,
\quad \mbox{where } a(\eta) \equiv -\frac1{H\eta}\;.
\ee
The full metric consists of the background metric $\hat{g}_{\mu\nu}$ and the conformally rescaled graviton field $h_{\mu\nu}$
\be
g_{\mu\nu}(\eta,\vec{x}) \equiv a^2(\eta) 
\Bigl(\eta_{\mu\nu} + \kappa h_{\mu\nu}(\eta,\vec{x}) \Bigr)\;.
\ee
Here $\kappa^2 \equiv 16 \pi G$ is the loop counting parameter of quantum gravity. 
The graviton propagator was obtained by adding a non-de Sitter invariant gauge fixing term  \cite{tw,w02} to the invariant Lagrangian \eqref{Lag} 
\bea \label{gauge}
\mathcal{L}_{\rm GF} & \equiv & -\frac12 \sqrt{-g} g^{\mu\nu} F_{\mu} F_{\nu} \; , \\
F_{\mu} & \equiv & 
\eta^{\rho\sigma} \Bigl( h_{\mu\rho , \sigma} - \frac12 h_{\rho \sigma , \mu} 
+ (D-2) a H h_{\mu \rho} \delta^0_{\sigma} \Bigr)\;. 
\eea

Owing to the conformal coupling term, the matter sector of the bare Lagrangian \eqref{Lag} is invariant 
\bea
\tilde{\mathcal{L}}_{\rm{Matter}} = 
-\frac12 \partial_{\mu} \tilde{\phi} \partial_{\nu} \tilde{\phi} \tilde{g}^{\mu\nu} \sqrt{-\tilde{g}} 
- \frac{D-2}{8(D-1)} \tilde{\phi}^2 \tilde{R}\sqrt{-\tilde{g}} 
= -\frac12 \partial_{\mu} \phi \partial_{\nu} \phi g^{\mu\nu} \sqrt{-g} 
- \frac{D-2}{8(D-1)} \phi^2 R\sqrt{-g}
= \mathcal{L}_{\rm{Matter}} \;.
\eea
under the conformal rescaling
\bea
\tilde{g}_{\mu\nu} \;\equiv\; \Omega^{-2} g_{\mu\nu}
\;\;{\rm{and}}\;\;
\tilde{\phi} \;\equiv\; \Omega^{\frac{D-2}{2}} \phi\;.
\eea
Taking $\Omega = a$ gives the conformally rescaled metric as 
\be
\tilde{g}_{\mu\nu} = \eta_{\mu\nu} + \kappa h_{\mu\nu}\;.
\ee
We obtained the linearized effective CC scalar field equation with this conformally rescaled metric 
\bea\label{lineffscfe0}
\partial_{\mu} \Bigl( \sqrt{-\tilde{g}} \; \tilde{g}^{\mu\nu} \; \partial_{\nu}  \Bigr) \;\tilde{\phi}(x)
- \frac16  \tilde{\phi}(x) \; \tilde{R} \; 
- \int d^4x' M^2(x;x') \; \tilde{\phi}(x') \; 
= 0 \;.
\eea
The scalar self-mass-squared $M^2(x;x')$ was computed and renormalized 
by subtracting off the one loop divergences using four counterterms  \cite{bkp14, bkp17} 
\bea
\Delta \tilde{\mathcal{L}} = \frac12 c_1 \kappa^2 \square \tilde{\phi} \square \tilde{\phi} a^2 
-\frac12 c_2 \kappa^2 H^2 \partial_{\mu} \tilde{\phi} \partial^{\mu} \tilde{\phi}
+\frac12 c_3 \kappa^2 H^4 \tilde{\phi}^2 a^2
-\frac12 c_4 \kappa^2 H^2 \nabla \tilde{\phi} \cdot \nabla \tilde{\phi}\;.
\label{L_counter}
\eea
Each coefficient $c_i$ has two parts: a divergent part $-d_i$ which is chosen to exactly cancel the divergences occurring in the self-mass-squared, and a finite part $\Delta c_i$ which remains arbitrary.  
The fully renormalized self-mass-squared takes the form \cite{bkp14, bkp17} 
\bea
-iM^2(x;x') &=& i \kappa^2 a^2 \Bigl( \Delta c_1 \square^2 + \Delta c_2 H^2 \square + \Delta c_3 H^4 + \Delta c_4 \frac{H^2}{a^2}\nabla^2 + 
4\Delta c_1\frac{H}{a} \partial_{0} \square \Bigr) \delta^4(x-x')
\nonumber\\
& &+ {\rm Table ~\ref{cct4}} + {\rm Table ~\ref{kt4}}  + {\rm Table ~\ref{ct4}} 
\nonumber\\
& &+ i \kappa^2 \frac{1}{(4\pi)^2}\; a^2 \Bigl( \frac{11}{4} \ln(aa') H^2 \square 
+ [\frac{9}{40} + \frac{11}{2} \ln(aa') ]  \frac{H^3}{a} \partial_{0} 
\nonumber\\
& &\hspace{2.4cm}
+ [2- 6 \ln(aa')] H^4 - \frac32 \ln(aa') \frac{H^2}{a^2} \nabla^2\Bigr) \delta^4 (x-x')\;. 
\eea
Tables \ref{cct4}, \ref{kt4} and \ref{ct4} are given in Appendix \ref{app-nonlocal-tables}.
The Laplacian $\nabla^2$ and the d'Alembertian $\square$ are
\bea
\nabla^2 \equiv \partial_i \partial_i \qquad {\rm and} \qquad
\square \equiv  \frac1{a^2} [ -\partial_0^2 \;-\; 2 a H \partial_0 \;+\; \nabla^2 ]\; .
\eea
The de Sitter invariant length function is defined as 
\bea\label{ydef}
y(x;x') \equiv a(\eta) a(\eta') H^2 \Bigl\{ \Vert \vec{x} - \vec{x}' \Vert^2 -
(\vert \eta - \eta'\vert - i \delta)^2 \Bigr\} \;. 
\eea

\section{Effective Mode Equation}\label{effmodeeq}

In this section we solve the effective field equation \eqref{lineffscfe0} for the scalar mode functions. 
We begin by explaining the necessity of the Schwinger-Keldysh formalism for cosmological settings,   
then we discuss how to solve the effective field equation perturbatively. Finally we work out the one loop contributions to the CC scalar mode functions from inflationary gravitons.
\subsection{Schwinger-Keldysh formalism}\label{S-Kformalism}

The effective field equation \eqref{lineffscfe0} reveals two disturbing features if we were to interpret it in the sense of the \textit{in-out} formalism:
\begin{itemize}
\item{\textit{Causality violation} - the in-out effective field equation at a point $x^{\mu}$ receives influence from points $x'^{\mu}$ which lie outside the past light-cone of $x^{\mu}$; and}
\item{\textit{Reality violation} - the quantum-induced scalar field acquires an imaginary part due to the nonzero imaginary part of the in-out self-mass-squared.}
\end{itemize} 
Neither of these features means the in-out formalism is wrong. It is 
the right answer to questions about matrix elements between scattering states if the system begins in free vacuum in the asymptotic past and ends up the same way in the far future. However it is not very relevant for cosmology in which we do not know how the universe begins and ends. The question of greater relevance for cosmology is how the system evolves if released at finite time in a prepared state. The Schwinger-Keldysh (or \textit{in-in}) formalism \cite{Schwinger, Mahanthappa, Bakshi, Keldysh} 
answers this more relevant question and it is almost as simple to use as the Feynman diagrams of the in-out formalism. Therefore we would like to solve the linearized effective field equation for the plane wave mode solution  $\tilde{\Phi}(x;\vec{k})$ 
which is given in terms of $\tilde{\phi}(x)$ as 
\bea
\tilde{\Phi}(x;\vec{k}) = \Bigl\langle \Omega_{\rm in} \Bigl\vert \Bigl[\tilde{\phi}(x),
\alpha^{\dagger}(\vec{k}) \Bigr] \Bigr\vert \Omega_{\rm in} \Bigr\rangle \; ,\label{SKeqn}
\eea
where  $\alpha^{\dagger}(\vec{k}) $ is the creation operator and $\Bigr\vert \Omega_{\rm in} \Bigr\rangle $ is the in-vacuum state.
Because excellent reviews on the Schwinger-Keldysh formalism exist \cite{Chou, Jordan, Calzetta, Ford},
we merely comment how to use it. 

The distinct property of the Schwinger-Keldysh formalism compared to the in-out formalism is that each end of particle lines carries a $\pm$ polarity. 
Therefore, every in-out propagator 
gives rise to four Schwinger-Keldysh propagators. In our case, the in-out propagators depend on the de Sitter invariant length function $y(x;x')$ and 
the four propagators can be obtained by the following substitutions for $y(x;x')$  
\bea
i\Delta_{\scriptscriptstyle ++}(x;x'): \quad
y &\rightarrow&
y_{\scriptscriptstyle ++}(x;x) \equiv H^2 a a' \Bigl[ \Vert \vec{x}
\!-\! \vec{x}' \Vert^2 - (\vert \eta \!-\! \eta' \vert \!-\! i\delta)^2 \Bigr] \; , \qquad 
\label{y++}
\\
i\Delta_{\scriptscriptstyle +-}(x;x') : \quad
y &\rightarrow&
y_{\scriptscriptstyle +-}(x;x) \equiv H^2 a a' \Bigl[ \Vert \vec{x}
\!-\! \vec{x}' \Vert^2 - (\eta \!-\! \eta' \!+\! i\delta)^2 \Bigr] \; , \qquad 
\label{y+-}
\\
i\Delta_{\scriptscriptstyle -+}(x;x') : \quad
y &\rightarrow&
y_{\scriptscriptstyle -+}(x;x) \equiv H^2 a a' \Bigl[ \Vert \vec{x}
\!-\! \vec{x}' \Vert^2 - (\eta \!-\! \eta' \!-\! i\delta)^2 \Bigr] \; , \qquad 
\label{y-+}
\\
i\Delta_{\scriptscriptstyle --}(x;x') : \quad
y &\rightarrow&
y_{\scriptscriptstyle --}(x;x) \equiv H^2 a a' \Bigl[ \Vert \vec{x}
\!-\! \vec{x}' \Vert^2 - (\vert \eta \!-\! \eta' \vert \!+\! i\delta)^2 \Bigr] \; . \qquad
\label{y--}
\eea
Because external lines can be either $+$ or $-$, every 1PI $N$-point function of the in-out formalism gives rise to $2^N$  Schwinger-Keldysh 1PI $N$-point functions. The Schwinger-Keldysh effective action generates these 1PI functions. For our case of the 1PI 2-point function for the CC scalar field, it derives from
\bea
\lefteqn{\Gamma[\tilde{\phi}_{+}, \tilde{\phi}_{-}] = S[\tilde{\phi}_{+}] - S[\tilde{\phi}_{-}] -\frac{1}{2}\int d^4x\int d^4x'}
\nn \\ 
&&\biggl\{ \tilde{\phi}_{+}(x)M^2_{++}(x;x') \tilde{\phi}_{+}(x') +  \tilde{\phi}_{+}(x)M^2_{+-}(x;x') \tilde{\phi}_{-}(x')
+  \tilde{\phi}_{-}(x)M^2_{-+}(x;x') \tilde{\phi}_{+}(x') +  \tilde{\phi}_{-}(x)M^2_{--}(x;x') \tilde{\phi}_{-}(x')\biggr\}
\nn \\ 
&& + \mathcal{O}(\tilde{\phi}^{3}_{\pm})\;.
\eea
Here $\tilde{\phi}_{+}$ carries out forward time evolution from the prepared state 
while $\tilde{\phi}_{-}$ carries out backward evolution to the original state, and $S[\tilde{\phi}]$ is the classical action.   
Varying this action with respect to either $\tilde{\phi}_{+}$ or $\tilde{\phi}_{-}$
and setting them equal to $\tilde{\phi}$, we have the Schwinger-Keldysh effective field equation, which is the effective field equation given in \eqref{lineffscfe0} with $M^2(x;x')$ replaced by $M^2_{\scriptscriptstyle ++}(x;x') + M^2_{\scriptscriptstyle +-}(x;x')$, 
\bea
\partial_{\mu} \Bigl( \sqrt{-\tilde{g}} \; \tilde{g}^{\mu\nu} \; \partial_{\nu}  \Bigr) \;\tilde{\Phi}(x;\vec{k})
- \frac16  \tilde{\Phi}(x;\vec{k}) \; \tilde{R} \; 
- \int_{\eta_i}^0 \! d\eta'\!\int\!d^3x' \,\Bigl\{M^2_{\scriptscriptstyle ++}(x;x') + M^2_{\scriptscriptstyle +-}(x;x')\Bigr\} \tilde{\Phi}(x';\vec{k}) \; 
= 0 \;.
\label{linear_eq_ren}
\eea
Here $\eta_i = -1/H$ is the initial time at which the universe is assumed to be in free vacuum (i.e., in the Bunch-Davies vacuum). 
At one loop order $i M^2_{++}(x;x')$ is identical to the in-out self-mass-squared  $i M^2(x;x')$.
The $+-$ self-mass-squared $i M^2_{+-}(x;x')$ can be obtained by the substitution rule given in \eqref{y+-}. One can check by examining the relations \eqref{y++} and \eqref{y+-}, the retarded self-mass-squared (the bracketed term in \eqref{linear_eq_ren}) vanishes for $\eta' > \eta$ and it is real for $\eta > \eta'$. That is, the Schwinger-Keldysh effective field equation \eqref{linear_eq_ren} is causal and real as desired.

\subsection{Perturbative Solution}\label{persol}

Because we only have one loop result for the scalar self-mass-squared, we must solve \eqref{linear_eq_ren} perturbatively by expanding it in powers of the loop-counting parameter $\kappa^2 \equiv 16\pi G$: 
\bea
M_{\rm ret}^2(x;x') \equiv M^2_{\scriptscriptstyle ++}(x;x') + M^2_{\scriptscriptstyle +-}(x;x')
&=& \sum_{\ell=1}^{\infty} \kappa^{2\ell} \mathcal{M}^2_{\ell}(x;x')\;, 
\label{Mret-expansion}
\\
\tilde{\Phi}(x;\vec{k})
&=& \sum_{\ell=0}^{\infty} \kappa^{2 \ell} \tilde{\Phi}_{\ell}(x;\vec{k})  
= \sum_{\ell=0}^{\infty} \kappa^{2 \ell} \tilde{u}_{\ell}(\eta,k) \times  e^{i\vec{k}\cdot\vec{x}}\;,
\label{phi-expansion}
\eea
where $\tilde{u}_{\ell}(\eta, k)$ are plane wave mode functions.

By substituting these expansions to the effective field equation \eqref{linear_eq_ren}, the zeroth order solution is\footnote{Since we are working with the conformally rescaled metric $\tilde{g}_{\mu\nu} = \eta_{\mu\nu} + \kappa h_{\mu\nu}$, the zeroth order values are the same as in flat space, i.e., $\tilde{R}_{0} = 0$.}
\bea
 \tilde{\Phi}_0(x;\vec{k}) = \tilde{u}_{0}(\eta, k)  e^{i\vec{k}\cdot\vec{x}} =  \frac{e^{-ik\eta}}{\sqrt{2k}} e^{i\vec{k}\cdot\vec{x}} \;.
\eea
The first order equation
\bea\label{lineffmodeeqn}
\partial^{2} \; \tilde{\Phi}_{1}(x;\vec{k}) 
- \int_{\eta_i}^0 d\eta' \int d^3x' \mathcal{M}^2_{1}(x;x') \; \tilde{\Phi}_{0}(x';\vec{k}) \;= 0\;,
\label{linear_eq_ren-pert}
\eea
and its solution takes the same form
\be
\tilde{\Phi}_{1}(x;\vec{k}) = \tilde{u}_1(\eta, k)  e^{i\vec{k}\cdot\vec{x}}\;,
\ee 
where the one loop corrected mode function $\tilde{u}_1(\eta, k)$ 
can be obtained by solving the following mode equation
\be
( -\partial_{\eta}^2 -k^2) \tilde{u}_{1}(\eta, k) 
= \int_{\eta_i}^0 \! d\eta' \tilde{u}_{0}(\eta', k) \!\int\!d^3x'  \mathcal{M}^2_{1}(x;x') 
e^{i \vec{k}\cdot (\vec{x}' - \vec{x})}  \; . \qquad 
\label{linear_eq_ren-1st}
\ee

The remaining task is then performing the integration of the right hand side of \eqref{linear_eq_ren-1st}. One fact to keep in mind is that we have taken the initial state to be the Bunch-Davies vacuum, but we have not worked out loop corrections to the initial vacuum. This means that the first order solution $\tilde{u}_1(\eta, k)$ only makes sense at late times  $\eta \rightarrow 0^-$.  To see the leading late time behavior of $\tilde{u}_1$, it is illuminating to convert \eqref{linear_eq_ren-1st} from conformal time $\eta$ to physical time $t \equiv -\ln(-H \eta)/H$,
\bea
\label{keyeqn}
\Bigl[ \frac{\partial^2}{\partial t^2} + 3H \frac{\partial}{\partial t} + \frac{k^2}{a^2}\Bigr] \tilde{u}_1(t,k) 
= -\frac1{a^2} \int_{\eta_i}^0 d\eta'  \tilde{u}_0(\eta',k) \int  d^3x' \mathcal{M}^2_1(x;x') \; e^{i \vec{k}
\cdot (\vec{x}' - \vec{x})}\;.
\eea
Given a presumed form for the late time limit of the right hand side (r.h.s.), the late time behavior of $\tilde{u}_1$ can be easily inferred as follows:
\bea
{\rm r.h.s.} \longrightarrow  a \quad &\Longrightarrow& \tilde{u}_1
\longrightarrow {\rm Constant} + \frac{a}{4 H^2} \; ,\label{s2strl} \\
{\rm r.h.s.} \longrightarrow 1  \quad &\Longrightarrow& \tilde{u}_1 
\longrightarrow {\rm Constant} + \frac{\ln(a)}{3H^2} \;,\label{2ndrl} \\
{\rm r.h.s.} \longrightarrow \frac1{a} \quad &\Longrightarrow& \tilde{u}_1
\longrightarrow {\rm Constant} -\frac{1}{2H^2 a} \; ,\label{4thrl} \\
{\rm r.h.s.} \longrightarrow a \ln(a) \quad &\Longrightarrow&  \tilde{u}_1
\longrightarrow {\rm Constant} + \frac{a\ln(a)}{4 H^2} - \frac{a}{4H^2}\;,\label{s1strl} \\
{\rm r.h.s.} \longrightarrow \ln(a)  \quad &\Longrightarrow& \tilde{u}_1
\longrightarrow {\rm Constant} +\frac{\ln^2(a)}{6H^2} -\frac{\ln(a)}{9H^2} \; ,\label{1strl} \\
{\rm r.h.s.} \longrightarrow \frac{\ln(a)}{a} \quad &\Longrightarrow& \tilde{u}_1
\longrightarrow {\rm Constant} -\frac{\ln(a)}{2H^2a} -\frac{1}{4H^2 a}\;. \label{3rdrl}
\eea

Collecting all the contributions from conformal-conformal \cite{bkp14}, kinetic-kinetic \cite{bkp17} and kinetic-conformal cross \cite{bkp17} parts, the regularized one loop self-mass-squared is 
\bea\label{Totalregkccc}
-iM^2_{\rm reg}(x;x') &=& -iM^2_{\rm regconf}(x;x') \;+\; -iM^2_{\rm regK}(x;x') \;+\; -iM^2_{\rm regcross}(x;x') \;,
\nonumber\\
&=& i \kappa^2 a^2 \Bigl( d_1 \square^2 + d_2 H^2 \square + d_3 H^4 + d_4 \frac{H^2}{a^2} \nabla^2 + d_5 \frac{H}{a} \partial_{0} \square\Bigr) \delta^D(x-x') 
\nonumber\\
& &+ {\rm Table ~\ref{cct4}} + {\rm Table ~\ref{kt4}}  + {\rm Table ~\ref{ct4}} 
\nonumber\\
& &+ i \kappa^2 \frac{1}{(4\pi)^2}\; a^2 \Bigl( \frac{11}{4} \ln(aa') H^2 \square 
+ [\frac{9}{40} + \frac{11}{2} \ln(aa') ]  \frac{H^3}{a} \partial_{0} 
\nonumber\\
& &\hspace{2.4cm}+ [2- 6 \ln(aa')] H^4 - \frac32 \ln(aa') \frac{H^2}{a^2} \nabla^2\Bigr) \delta^4 (x-x') + \mathcal{O}(D-4)\;,
\eea
where Tables  \ref{cct4}, \ref{kt4} and \ref{ct4} are given in Appendix \ref{app-nonlocal-tables} and the coefficients $d_i$ are from combining results of \cite{bkp14, bkp17},
\bea
d_1 
&=& \frac{H^{D-4}}{(4\pi)^\frac{D}{2}} \Biggl\{-\frac{(D-2)}{2(D-4)(D-3)(D-1)} + \frac16\gamma +\mathcal{O}(D-4)\Biggr\}\;,\label{d1-final} \\
d_2 
&=& \frac{H^{D-4}}{(4 \pi)^{\frac{D}{2}}} \Biggr\{\frac{8D^8-127D^7+782D^6 -2465D^5 +4386D^4 -4536D^3 +2848D^2 -1016D +192}{2^4(D-4)(D-3)(D-1)^3}
\nonumber\\
& & \hspace{1.5cm}- \frac{327}{80}  - \frac{98}{9} \gamma  + \mathcal{O}(D-4) \Biggr\}\;, \\
d_3 
&=& \frac{H^{D-4}}{(4 \pi )^{\frac{D}{2}}} 
\Biggl\{ \frac{45D^6 -650D^5 + 3435D^4 -8520D^3 +10114D^2 -4456D -178}{2^5 5 (D-4)(D-3)(D-1)} 
\nonumber\\
& & \hspace{1.5cm}-\frac{33421}{3600} + \frac{2163}{80} \gamma  +  \mathcal{O}(D-4) \Biggr\}\;, 
\eea
\bea
d_4 
&=& \frac{H^{D-4}}{(4 \pi)^{\frac{D}{2}}} \Biggl\{
-\frac{36D^7-459D^6+2070D^5-7695D^4+14290D^3+316D^2-20804D+7440}{2^3 3^2(D-4)(D-3)(D-1)^2}
\nonumber\\
& & \hspace{1.5cm} +\frac{2759}{540} - \frac{2327}{9} \gamma  + \mathcal{O}(D-4) \Biggr\}\;, \\
d_5 
&=& \frac{H^{D-4}}{(4\pi)^\frac{D}{2}} \Biggl\{-\frac{(D-2)^2}{(D-4)(D-3)(D-1)} + \frac23 \gamma +  \mathcal{O}(D-4) \Biggr\}\;.\label{d5-final}
\eea
Here $\gamma \approx 0.577215$ is Euler's constant.

Choosing the coefficients $c_i$ of the counterterms
\bea
c_i &=& -d_i + \Delta c_i \quad \mbox{for }  i= 1, 3, 4\;, \\
c_2 &=& -d_2 + (D-2)d_1 + \Delta c_2 \;, 
\eea
where $\Delta c_i$ are arbitrary finite terms that remain after cancelling the divergent parts occurring in the primitive diagrams, and exploiting the relation between $d_1$ and $d_5$
\be
d_5 = 2(D-2)d_1 \;,
\ee
lead to the cancellation of all the divergent terms.
Finally taking the unregulated limit ($D=4$) gives 
\bea\label{Totalrenkccc}
-iM^2_{\rm ren}(x;x') &=& i \kappa^2 a^2 \Bigl( \Delta c_1 \square^2 + \Delta c_2 H^2 \square + \Delta c_3 H^4 + \Delta c_4 \frac{H^2}{a^2}\nabla^2 +
4\Delta c_1\frac{H}{a} \partial_{0} \square \Bigr) \delta^4(x-x')
\nonumber\\
& &+ {\rm Table ~\ref{cct4}} + {\rm Table ~\ref{kt4}}  +  {\rm Table ~\ref{ct4}} 
\nonumber\\
& &+ i \kappa^2 \frac{1}{(4\pi)^2}\; a^2 \Bigl( \frac{11}{4} \ln(aa') H^2 \square 
+ [\frac{9}{40} + \frac{11}{2} \ln(aa') ]  \frac{H^3}{a} \partial_{0} 
\nonumber\\
& &\hspace{2.4cm}+ [2- 6 \ln(aa')] H^4 - \frac32 \ln(aa') \frac{H^2}{a^2} \nabla^2\Bigr) \delta^4 (x-x') \;.
\eea
\subsubsection{Local Corrections}\label{lsemf}

By inserting the local terms proportional to $\delta^4(x-x')$ into the right hand side of Eq. \eqref{keyeqn} and performing the integration gives local corrections to $\tilde{u}_1$ at late times ($\eta \rightarrow 0^{-}$ or equivalently $a \rightarrow \infty$), 
\bea
\label{ls}
\tilde{u}_{\rm local} &=& \tilde{u}_{0} + \kappa^2 \tilde{u}_{1 {\rm local}} + \mathcal{O}(\kappa^4)\;,
\nn\\
&\sim& \frac{1}{\sqrt{2k}} \biggl\{1 - \frac{\kappa^2 H^2}{2^4 \pi^2} \; \Bigl(\Delta c_4 - \frac34 \Bigr) \; \ln(a) \biggr\} \;.
\eea
While the zeroth order mode function $\tilde{u}_0$ goes to a constant, 
the first order local contribution to the mode function $\tilde{u}_{1{\rm local}}$ 
grows like $\ln(a)$ though it is suppressed by a factor of $\kappa^2H^2$.  
\subsubsection{Nonlocal Corrections}\label{nlc}

Nonlocal corrections come from inserting Tables \ref{cct4} - \ref{ct4} of nonlocal terms in Appendix \ref{app-nonlocal-tables}
into Eq. \eqref{keyeqn}. We apply a seven-step procedure developed in \cite{kw08} to evaluate \eqref{keyeqn}:
\begin{itemize}
\item{{\it Step 1}: 
Convert any factors of $1/y$ to $\ln(1/y)$ using the identities,
\bea
\frac4{y} 
&=& \frac{\square}{H^2} \Biggl\{\ln\Bigl(\frac{y}4\Bigr)\Biggr\} + 3 \;, \label{1/x} 
\\
\frac4{y} \ln\Bigl(\frac{y}4\Bigr) 
&=& \frac{\square}{H^2} \Biggl\{ \frac12 \ln^2\Bigl(\frac{y}4\Bigr) -
\ln\Bigl(\frac{y}4\Bigr) \Biggr\} + 3\ln\Bigl(\frac{y}4\Bigr) - 2 \;. \label{lnx/x}
\eea}\label{step1}
\item{{\it Step 2}: 
Factor out $\square$ and $\nabla^2$ (which are w.r.t. $x$) outside the integration over $x^{\prime \mu}$ using the identities,
\bea
\frac{\square}{H^2} & \longrightarrow & 
-\Bigl[ a^2 \partial_{a}^2 + 4 a \partial_{a} + \frac{k^2}{a^2 H^2} \Bigr]\;, \label{box} \\
\nabla^2 & \longrightarrow & - k^2\;.
\eea}\label{step2}
\item{{\it Step 3}:
Combine the $++$ and $+-$ terms to extract a factor of $i$ and a step function $\theta$, which makes the effective mode equation \eqref{keyeqn} manifestly real and causal,
\bea
\ln\Bigl(\frac{y_{\scriptscriptstyle ++}}4\Bigr) -
\ln\Bigl(\frac{y_{\scriptscriptstyle +-}}4\Bigr) 
&=& 2 \pi i \theta \Bigl( \Delta \eta  - \Delta x \Bigr)\;, 
\\
\ln^2 \Bigl( \frac{y_{\scriptscriptstyle ++}}4 \Bigr) 
- \ln^2 \Bigl( \frac{y_{\scriptscriptstyle +-}}4 \Bigr) 
&=& 4 \pi i \theta\Bigl(\Delta \eta - \Delta x \Bigr) 
\ln \Bigl(\frac14 H^2 a a' (\Delta \eta^2 - \Delta x^2) \Bigr)\;. 
\eea
Here $a \equiv a(\eta)$, $a' \equiv a(\eta')$,
$\Delta \eta \equiv \eta - \eta'$ and $\Delta x \equiv \Vert \vec{x}
- \vec{x}'\Vert$. Note that the invariant length function $y$ becomes
\bea
y_{\scriptscriptstyle +\pm} \longrightarrow -\frac14  H^2 a a'
(\Delta \eta^2 - \Delta x^2)\;.
\eea}\label{step3}
\item{{\it Step 4}: 
Make the change of variables $\vec{r} = \vec{x}' - \vec{x}$
to perform the angular integrations, and make the change of variables $r = \Delta \eta \cdot z$,
\bea
\int  d^3x' \theta\Bigl(\Delta \eta - \Delta x\Bigr) &F& \Bigl(\frac14 H^2 a a'(\Delta \eta^2 - \Delta x^2)\Bigr)
\;e^{i \vec{k} \cdot (\vec{x}' - \vec{x})} 
\nonumber\\
& = & 4\pi \theta(\Delta \eta) \int_0^{\Delta \eta} dr \, r^2 F \Bigl(\frac14 H^2 a a' (\Delta \eta^2 - r^2)\Bigr)
\frac{\sin(k \Delta x)}{k \Delta x}\;, 
\\
& = & 4\pi \theta(\Delta \eta) \Delta \eta^3 \int_0^1  dz \, z^2 F\Biggl(a a' \Bigl(\frac1{a'} - \frac1{a}\Bigr)^2 
\Bigl(\frac{1 - z^2}4\Bigr) \Biggr) \frac{\sin(k \Delta \eta z)}{k \Delta \eta z}\;. \label{ang} 
\eea}\label{step4}
\item{{\it Step 5}:
Perform the integration over $z$, which leads to a combination of elementary
functions and sine and cosine integrals \cite{dw}.}
\label{step5}
\item{{\it Step 6}:
Make the last change of variables $a' = -1/ H\eta'$ and perform the integration over $a'$.}\label{step6}
\item{{\it Step 7}:
Act external derivatives on the result of the integration with respect to $a$ using \eqref{box}.}\label{step7}
\end{itemize}

Also, we adopt a simplification made in Ref. \cite{kw08}, which is to take the zeroth order solution $\tilde{u}_0(\eta, k)$ as its late time value $\tilde{u}_0(0,k) = 1/\sqrt{2k}$ and ignore the spatial phase factor $e^{i\vec{k}\cdot (\vec{x}'-\vec{x})}$:
\bea\label{replace}
-\frac1{a^2} \int_{\eta_i}^{0} \; d\eta' \; \int \; d^3x'
\mathcal{M}^2_1(x;x') \tilde{u}_0(\eta',k) e^{i \vec{k} \cdot (\vec{x}' -
\vec{x})}  \longrightarrow - \frac{\tilde{u}_0(0,k)}{a^2} \; \int_{\eta_i}^{0}
\; d\eta' \; \int \; d^3x' \mathcal{M}^2_1(x;x')\;.
\eea
This results in only elementary functions in the $z$ integration (without sine and cosine integrals).
Also, this simplification grants us neglecting any contribution from Tables  \ref{cct4}, \ref{kt4}, and \ref{ct4} which contains a factor of $\nabla^2$. Then the integrals of the nonlocal terms take the following form,
\bea\label{fornewtables}
-\tilde{u}_0(0,k) \frac{i H^8 a^{4-K}}{(4 \pi)^4}
\Bigl(\frac{\square}{H^2}\Bigr)^N  \int_{\eta_i}^{0} 
d\eta'  a^{\prime K} \int  d^3x' 
\Biggl\{
f\Bigl(\frac{y_{\scriptscriptstyle ++}}4\Bigr) -
f\Bigl(\frac{y_{\scriptscriptstyle +-}}4\Bigr)
\Biggr\}\;, 
\eea
where the constant $K$ takes the values of 1, 2, 3, and 4 and the functions $f(x)$ are $1/x$, $\ln(x)/x$, $\ln{x}$ and $\ln^2{x}$. 
The results of the integrations are given in Tables \ref{ia1}-\ref{ia4} in Appendix \ref{app-key-integrals}, and one example calculation using the seven-step procedure is provided in Appendix \ref{app-example-key-integral}. 
The results of \eqref{fornewtables} from conformal, kinetic  and cross nonlocal parts are given in Tables \ref{newcct4}, \ref{newkt4} and \ref{newct4} in Appendix \ref{app-result-tables}.  

Finally, summing the results of Tables \ref{newcct4}, \ref{newkt4} and \ref{newct4} (which are in the form of 
$\#_1a\ln(a) + \#_2 a$) and using \eqref{s2strl} and \eqref{s1strl} gives
\bea
\label{nls}
\tilde{u}_{\rm nonlocal} &=& \tilde{u}_{0} + \kappa^2 \tilde{u}_{1{\rm nonlocal}} + \mathcal{O}(\kappa^4)\;,
\nn\\
&\sim& \frac{1}{\sqrt{2k}} \biggl\{ 1 - \frac{\kappa^2 H^2}{\pi^2}\Bigl[-\frac{3603}{2^9 \cdot 15} \; a\ln(a) \;+\; \frac{50033}{2^{10} \cdot 15} \; a\Bigr] \biggr\} \;.
\eea

The first order nonlocal contribution to the mode function $\tilde{u}_{1{\rm nonlocal}}$ grows like $a\ln(a)$. It is dominant over the local contribution of $\ln(a)$ in \eqref{ls} at late times ($a \rightarrow \infty$), thus the arbitrariness $\Delta c_4$ from the local counterterm is negligible in the late time limit.
\section{Discussion}\label{dis}

We have solved the one loop corrected, Schwinger-Keldysh effective field equation for a CC scalar interacting with a graviton during de Sitter inflation. It is instructive to take the inverse conformal transformation
\be
\phi = \frac{\tilde{\phi}}{a}\; \mbox{ and }\; u = \frac{\tilde{u}}{a}\;,
\ee
and express the result in physical time $t$,
\bea
u_{\rm CC} 
\sim 
\frac{1}{\sqrt{2k}} \biggl\{\frac{1}{a} + GH^2\Bigl[\frac{3603}{2^5 \cdot 15 \pi} \; \ln(a) \;-\; \frac{50033}{2^6 \cdot 15 \pi} \;  -\Bigl(\Delta c_4 - \frac34 \Bigr) \; \frac{\ln(a)}{a} \Bigr] + \mathcal{O}(G^2H^4) \biggr\} \;.
\eea   
While the tree order CC scalar mode functions redshift to zero at late times, the one corrected mode functions gain a secular growth effect via the interaction with gravitons.
This is in contrast to the case of the MMC scalar, for which the tree order mode functions become constant at late times, but they do not get significant corrections at one loop order \cite{kw07, kw08},  
\bea
u_{\rm MMC} 
\sim 
\frac{1}{\sqrt{2k}} \biggl\{\frac{H}{k} + \mathcal{O}(G^2H^4) \biggr\} \;.
\eea  
It is interesting to see that 
the inflationary production of CC scalars is suppressed by $1/a$ due to the conformal invariance, but their interactions with gravitons turn out to be significant because the scalar is not differentiated (so not redshifted) in the conformal coupling.
In the case of MMC scalar $+$ graviton, although they both are copiously produced during inflation, they only interact through their kinetic energies, which redshift away at late times, hence they do not make significant loop corrections \cite{pw, LPPW, kw07, kw08}.

Considering the undifferentiated scalar seems to be the key to have logarithmic corrections,  
this sort of secular loop effects might also present for a more general non-minimal coupling (such as $\xi R \phi^2$ with an arbitrary number $\xi$) in the context of Higgs inflation \cite{SBB-1989, BS-2008, WP-2010}. 
On the other hand, even though the loop counting parameter $GH^2$ is extremely small, 
the secular factor 
$\ln(a)$ would grow and eventually overcome it, then perturbation theory would break down. Hence it would be necessary to use a nonperturbative resummation method such as Starobinsky's stochastic technique \cite{Starobinsky-stochastic}. There are nonperturbative techniques for various interactions in the literature \cite{Starobinsky-stochastic, Starobinsky-Yokoyama, W-2005, TW-2005, mw, kw06, ko, PTW-2007, Finelli:2008zg, Finelli:2010sh, Youssef-1301, Serreau-1302, Serreau-1611}.
However, we leave these considerations for future work.

\vskip 1cm
\centerline{\bf Acknowledgements}
\vskip 0.3cm

We thank Markus Fr\"ob and Richard P. Woodard for helpful comments and discussions. 
EOK acknowledges TUBA-GEBIP 2015 awards programme. SB and EOK acknowledge the support from Istanbul Technical University, The Scientific Research Project (ITU-BAP) coordination unity with Project Number: 39955.



\newpage
\appendix

\section{The Tables For All Finite Nonlocal Contributions From \cite{bkp14, bkp17}}\label{app-nonlocal-tables}

Table ~\ref{cct4}  uses $\tilde{\kappa}^2 \equiv (\frac{D-2}{8(D-1)})^2\kappa^2$ and Table ~\ref{ct4} uses $\tilde{\tilde{\kappa}}^2 \equiv \frac{D-2}{8(D-1)} \kappa^2$.
\begin{table}[htbp]
\begin{center}
\caption{All Finite Nonlocal ``conformal'' contributions with $x \equiv \frac{y}{4}$, where $y(x;x')$ is defined in the equation \eqref{ydef} \cite{bkp14}.}
\begin{ruledtabular}
\begin{tabular*}{1.5\textwidth}{ll}
$\phantom{ssssssssss}\rm External \; operators$&$\mbox{Coeff. of } \frac{\tilde{\kappa}^2 H^4}{(4\pi)^4}$\\[1ex]
\hline\\
$\phantom{ssssssssss}(aa')^3 \square^3/H^2$&$\frac{9\ln{x}}{5x}\phantom{ssssssssssssssssssssss}$\\[2ex]
$\phantom{ssssssssss}(aa')^3 \square^2$&$\frac{9\ln{x}}{x} - \frac{21}{5x}$\\[2ex]
$\phantom{ssssssssss}(aa')^3 H^2\square$&$-\frac{267\ln{x}}{5x} + \frac{51}{x}$\\[2ex]
$\phantom{ssssssssss}(aa')^3 H^4$&$\frac{258\ln{x}}{5x} - \frac{549}{5x}$\\[2ex]
$\phantom{ssssssssss}(aa')^2(a^2+a'^2)\square^3/H^2$&$-\frac{9\ln{x}}{10x}$\\[2ex]
$\phantom{ssssssssss}(aa')^2(a^2+a'^2)\square^2$&$-\frac{99\ln{x}}{10x} - \frac{9}{10x}$\\[2ex]
$\phantom{ssssssssss}(aa')^2(a^2+a'^2)H^2\square$&$\frac{36\ln{x}}{x} - \frac{51}{2x}$\\[2ex]
$\phantom{ssssssssss}(aa')^2(a^2+a'^2)H^4$&$-\frac{108\ln{x}}{5x} + \frac{174}{5x}$\\[2ex]
$\phantom{ssssssssss}(aa')^2 H^2 \nabla^2$&$\frac{128\ln{x}}{5x} + \frac{166}{5x}$\\[2ex]
$\phantom{ssssssssss}(aa')^2 \nabla^2 \square$&$-\frac{64\ln{x}}{5x}$\\[2ex]
$\phantom{ssssssssss}(aa')(a^2+a'^2) H^2 \nabla^2$&$-\frac{4\ln{x}}{x} - \frac{2}{x}$\\[2ex]
$\phantom{ssssssssss}(aa')(a^2+a'^2)\nabla^2\square$&$\frac{2\ln{x}}{x} + \frac{2}{x}$\\[2ex]
$\phantom{ssssssssss}(aa')\nabla^4$&$0$\\[2ex]
\end{tabular*}\label{cct4}
\end{ruledtabular}
\end{center}
\end{table}\begin{table}[htbp]
\begin{center}
\caption{All Finite Nonlocal ``kinetic'' contributions with $x \equiv \frac{y}{4}$, where $y(x;x')$ is defined in the equation (\ref{ydef}) \cite{bkp17}.}
\begin{ruledtabular}
\begin{tabular*}{1.5\textwidth}{ll}
$\phantom{ssssssssss}\rm External \; operators$&$\mbox{Coeff. of } \frac{\kappa^2 H^4}{(4\pi)^4}$\\[1ex]
\hline\\
$\phantom{ssssssssss}(aa')^3 \square^3/H^2$&$\frac{\ln{x}}{3x}\phantom{ssssssssssssssssssssss}$\\[2ex]
$\phantom{ssssssssss}(aa')^3 \square^2$&$-\frac{3\ln{x}}{2x} $\\[2ex]
$\phantom{ssssssssss}(aa')^3 H^2\square$&$\frac{25\ln{x}}{x}-\frac{3}{2x} $\\[2ex]
$\phantom{ssssssssss}(aa')^3 H^4$&$-\frac{44\ln{x}}{x} - \frac{26}{x} - \ln(aa') \frac{30}{x}$\\[2ex]
$\phantom{ssssssssss}(aa')^2(a^2+a'^2)\square^3/H^2$&$-\frac{\ln{x}}{6x}$\\[2ex]
$\phantom{ssssssssss}(aa')^2(a^2+a'^2)\square^2$&$\frac{7\ln{x}}{6x} - \frac{1}{6x}$\\[2ex]
$\phantom{ssssssssss}(aa')^2(a^2+a'^2)H^2\square$&$\frac{\ln{x}}{3x} + \frac{5}{6x}$\\[2ex]
$\phantom{ssssssssss}(aa')^2(a^2+a'^2)H^4$&$-\frac{4\ln{x}}{x} - \frac{7}{2x}$\\[2ex]
$\phantom{ssssssssss}(aa')^2 H^2 \nabla^2$&$\frac{4\ln{x}}{x} - \frac{14}{x} + \ln(aa')[9\ln(x) + \frac{20}{x}]$\\[2ex]
$\phantom{ssssssssss}(aa')^2 \nabla^2 \square$&$\frac{8\ln{x}}{3x} + \frac{1}{3x}$\\[2ex]
$\phantom{ssssssssss}(aa')^2 H^2\nabla^2 \square^2$&$\frac{\ln{x}}{3x} $\\[2ex]
$\phantom{ssssssssss}(aa')(a^2+a'^2) H^2 \nabla^2$&$\frac{22\ln{x}}{3x} - \frac{67}{6x} + \ln(aa') \frac{12}{x}$\\[2ex]
$\phantom{ssssssssss}(aa')(a^2+a'^2)\nabla^2\square$&$-\frac{11\ln{x}}{3x} + \frac{4}{6x}$\\[2ex]
$\phantom{ssssssssss}(aa')\nabla^4$&$-\ln(x) + \frac{16}{3x} + \ln(aa') [2\ln(x) - \frac{2}{x}]$\\[2ex]
\end{tabular*}\label{kt4}
\end{ruledtabular}
\end{center}
\end{table}
\begin{table}[htbp]
\begin{center}
\caption{All Finite Nonlocal ``cross'' contributions with $x \equiv \frac{y}{4}$, where $y(x;x')$ is defined in the equation (\ref{ydef}) \cite{bkp17}.}
\begin{ruledtabular}
\begin{tabular*}{1.5\textwidth}{ll}
$\phantom{ssssssssss}\rm External \; operators$&$\mbox{Coeff. of } \frac{\tilde{\tilde{\kappa}}^2 H^4}{(4\pi)^4}$\\[1ex]
\hline\\
$\phantom{ssssssssss}(aa')^3 \square^3/H^2$&$0\phantom{ssssssssssssssssssssss}$\\[2ex]
$\phantom{ssssssssss}(aa')^3 \square^2$&$-\frac{27\ln{x}}{x} $\\[2ex]
$\phantom{ssssssssss}(aa')^3 H^2\square$&$-\frac{378\ln{x}}{x} - \frac{59}{x}$\\[2ex]
$\phantom{ssssssssss}(aa')^3 H^4$&$\frac{608\ln{x}}{x} - \frac{436}{x}$\\[2ex]
$\phantom{ssssssssss}(aa')^2(a^2+a'^2)\square^3/H^2$&$\frac{\ln{x}}{x}$\\[2ex]
$\phantom{ssssssssss}(aa')^2(a^2+a'^2)\square^2$&$-\frac{28\ln{x}}{x} + \frac{1}{x}$\\[2ex]
$\phantom{ssssssssss}(aa')^2(a^2+a'^2)H^2\square$&$\frac{178\ln{x}}{x} - \frac{26}{x}$\\[2ex]
$\phantom{ssssssssss}(aa')^2(a^2+a'^2)H^4$&$-\frac{252\ln{x}}{x} - \frac{66}{x}$\\[2ex]
$\phantom{ssssssssss}(aa')^2 H^2 \nabla^2$&$-24\ln(x) - \frac{142\ln{x}}{3x} + \frac{508}{3x}$\\[2ex]
$\phantom{ssssssssss}(aa')^2 \nabla^2 \square$&$\frac{76\ln{x}}{3x}$\\[2ex]
$\phantom{ssssssssss}(aa')(a^2+a'^2) H^2 \nabla^2$&$-\frac{24\ln{x}}{x} + \frac{58}{x}$\\[2ex]
$\phantom{ssssssssss}(aa')(a^2+a'^2)\nabla^2\square$&$\frac{12\ln{x}}{x} + \frac{4}{x}$\\[2ex]
$\phantom{ssssssssss}(aa')\nabla^4$&$\frac{4}{x}$\\[2ex]
\end{tabular*}\label{ct4}
\end{ruledtabular}
\end{center}
\end{table}

\newpage

\section{Key Integral Tables}\label{app-key-integrals}

\begin{table}[htbp]
\begin{center}
\caption{Integrals with $a'$.}
\begin{ruledtabular}
\begin{tabular*}{1.5\textwidth}{ll}
$\phantom{ssssssssss}f(x) $&$ -\frac{i H^4}{16 \pi^2} \times \int d^4x' a^{\prime }
\{f(\frac{y_{++}}{4}) - f(\frac{y_{+-}}{4})\} $\\[1ex]
\hline\\
$\phantom{ssssssssss}\frac{1}{x}$&$-\frac{1}{2a} + \mathcal{O}(\frac{1}{a^2})\phantom{sssssssssssssssssssssssssssssssssss}$\\[2ex]
$\phantom{ssssssssss}\frac{\ln(x)}{x} $&$-\frac{\ln(a)}{2a}+\frac{5}{4a}+\mathcal{O} (\frac{\ln(a)}{a^2}) $\\[2ex]
$\phantom{ssssssssss}\ln(x)$&$-\frac{1}{4a}+\frac{1}{18}+\mathcal{O}(\frac{1}{a^2})$\\[2ex]
$\phantom{ssssssssss}\ln^2(x)$&$ \frac{\ln(a)}{9}-\frac{1}{3}-\frac{\ln(a)}{2a}+\frac{5}{4a}+\mathcal{O}(\frac{1}{a^2})$\\[2ex]
\end{tabular*}\label{ia1}
\end{ruledtabular}
\end{center}
\end{table}
\begin{table}[htbp]
\begin{center}
\caption{Integrals with $a'^2$.}
\begin{ruledtabular}
\begin{tabular*}{1.5\textwidth}{ll}
$\phantom{ssssssssss} f(x) $&$ -\frac{i H^4}{16 \pi^2}  \times \int d^4x' a^{\prime 2}
\{f(\frac{y_{++}}{4}) - f(\frac{y_{+-}}{4})\} $\\[1ex]
\hline\\
$\phantom{ssssssssss}\frac{1}{x}$&$ -\frac{1}{a}+\mathcal{O}(\frac{1}{a^2})\phantom{ssssssssssssssssssssssssssssss}$\\[2ex]
$\phantom{ssssssssss}\frac{\ln(x)}{x} $&$-\frac{\ln(a)}{a}+\frac{3}{a}+\mathcal{O}(\frac{\ln(a)}{a^2})$\\[2ex]
$\phantom{ssssssssss}\ln(x)$&$  \frac{1}{12} - \frac{1}{2a} + \mathcal{O}(\frac{1}{a^2})$\\[2ex]
$\phantom{ssssssssss}\ln^2(x)$&$ \frac{\ln(a)}{6}-\frac{19}{36}-\frac{\ln(a)}{a}+\frac{3}{a}+\mathcal{O}(\frac{\ln(a)}{a^2})$\\[2ex]
\end{tabular*}\label{ia2}
\end{ruledtabular}
\end{center}
\end{table}
\begin{table}[htbp]
\begin{center}
\caption{Integrals with $a'^3$.}
\begin{ruledtabular}
\begin{tabular*}{1.5\textwidth}{ll}
$\phantom{ssssssssss} f(x) $&$ -\frac{i H^4}{16 \pi^2}  \times \int d^4x' a^{\prime 3}
\{f(\frac{y_{++}}{4}) - f(\frac{y_{+-}}{4})\} $\\[1ex]
\hline\\
$\phantom{ssssssssss}\frac{1}{x}$&$-\frac{\ln(a)}{a}+\frac{1}{a}+\mathcal{O}(\frac{1}{a^2})\phantom{sssssssssssssssssssssssssss}$\\[2ex]
$\phantom{ssssssssss}\frac{\ln(x)}{x} $&$-\frac{\ln^2(a)}{2a}+\frac{2\ln(a)}{a}-\frac{3}{a}+\frac{\pi^2}{3a}+\mathcal{O}(\frac{\ln(a)}{a^2})$\\[2ex]
$\phantom{ssssssssss}\ln(x)$&$\frac16-\frac{\ln(a)}{2a}+\frac{1}{4a}+ \mathcal{O}(\frac{1}{a^2})$\\[2ex]
$\phantom{ssssssssss}\ln^2(x)$&$\frac13 \ln(a)-\frac{11}{9}-\frac{\ln^2(a)}{2a}+\frac{2\ln(a)}{a}-\frac{9}{4a}+\frac{\pi^2}{3a}+\mathcal{O} (\frac{\ln(a)}{a^2})$\\[2ex]
\end{tabular*}\label{ia3}
\end{ruledtabular}
\end{center}
\end{table}
\begin{table}[htbp]
\begin{center}
\caption{Integrals with $a'^4$.}
\begin{ruledtabular}
\begin{tabular*}{1.5\textwidth}{ll}
$\phantom{ssssssssss} f(x) $&$ -\frac{i H^4}{16 \pi^2} \times \int d^4x' a^{\prime 4}
\{f(\frac{y_{++}}{4}) - f(\frac{y_{+-}}{4})\} $\\[1ex]
\hline\\
$\phantom{ssssssssss}\frac{1}{x}$&$-\frac12 + \mathcal{O}(\frac{1}{a})\phantom{sssssssssssssssssssssssssssssssssss}$\\[2ex]
$\phantom{ssssssssss}\frac{\ln(x)}{x} $&$\frac34 + \mathcal{O} (\frac{1}{a})  $\\[2ex]
$\phantom{ssssssssss}\ln(x)$&$ \frac16 \ln(a) - \frac{11}{36} + \mathcal{O} (\frac{1}{a})  $\\[2ex]
$\phantom{ssssssssss}\ln^2(x)$&$\frac16 \ln^2 (a) - \frac89 \ln(a) + \frac74 - \frac{\pi^2}{9}  +  \mathcal{O} (\frac{\ln(a)}{a})  $\\[2ex]
\end{tabular*}\label{ia4}
\end{ruledtabular}
\end{center}
\end{table}

\section{The Examples For Calculating Key Integrals From Table ~\ref{ia3}}\label{app-example-key-integral}

As an example, let us choose the term $f(x) = \frac1{x}$ from the Table~ \ref{ia3} .
The seven-step procedure is applied as follows:
\bea
\int d^4 x' \; a^{'3} \; f(x') 
&=& \int d^4 x' \; a^{'3} \Biggl\{ f(\frac4{y_{++}}) - f(\frac4{y_{+-}}) \Biggr\}\;,  
\nonumber\\
&=& \int d^4 x' \; a^{'3} \Biggl\{ [\frac{\square }{H^2} \ln(\frac{y_{++}}{4}) + 3 ]\;-\; [\frac{\square }{H^2} \ln(\frac{y_{+-}}{4}) + 3 ] \Biggr\}\;,   
\quad \leftarrow \mbox{ Step 1}
\nonumber\\
&=& \frac{\square }{H^2} \int d^4 x' \; a^{'3} \Biggl\{ \ln(\frac{y_{++}}{4}) - \ln(\frac{y_{+-}}{4})\Biggr\}\;,   
\quad \leftarrow \mbox{ Step 2}
\nonumber\\
&=& \frac{\square }{H^2} \int d^4 x' \; a^{'3} \;2\pi i \; \theta (\Delta \eta - \Delta x )\;,
\quad \leftarrow \mbox{ Step 3}
\nonumber\\
&=& 8\pi^2 i \frac{\square }{H^2} \int_{\eta_i}^{\eta} d^4 \eta' a^{'3} \;(\Delta \eta)^3\; \int_0^1 dz \; z^2\;,
\quad \leftarrow \mbox{ Step 4 and 5}
\nonumber\\
&=& \frac{8\pi^2 i}{3H^4} \frac{\square }{H^2} \int_1^{a} d \;a' \; a^{'} \; (\frac1{a'}-\frac1{a})^3\;,
\quad \leftarrow \mbox{ Step 6}
\nonumber \\
&=& \frac{16\pi^2 i}{H^4} \;[-\frac{\ln(a)}{a}+\frac{1}{a}+\mathcal{O}(\frac{1}{a^2})]\;.
\quad \leftarrow \mbox{ Step 7}
\eea 

The same procedure can be used for  $f(x') = \ln(x)$ and $f(x') = \ln^2(x)$, but for the function $f(x') = \frac{\ln(x)}{x}$, a little more algebra is required. We choose our example again from Table~ \ref{ia3} to illustrate the algebra,
\bea
\int d^4 x' \; a^{'3} \; f(x') &=& \int d^4 x' \; a^{'3} \Biggl\{ f(\frac{\ln{x_{++}}}{x_{++}}) - f(\frac{\ln{x_{+-}}}{x_{+-}}) \Biggr\} \;,\;\;{\rm{taking}}\;\;f(x') \;=\; \frac{\ln{x}}{x}\;,
\nonumber\\
&=& \int d^4 x' \; a^{'3} \Biggl\{ \biggl( \frac{\square }{H^2} [\frac12 \ln^2 (\frac{y_{++}}{4}) - \ln (\frac{y_{++}}{4})] + 3 \ln (\frac{y_{++}}{4}) -2 \biggr)
\nonumber\\
& &\hspace{2.0cm}\;-\; \biggl( \frac{\square }{H^2} [\frac12 \ln^2 (\frac{y_{+-}}{4}) -\ln(\frac{y_{+-}}{4})] + 3\ln (\frac{y_{+-}}{4}) -2 \biggl)  
\Biggr\}\;,
\nonumber\\
&=& \frac{\square }{H^2}  \int d^4 x' \; a^{'3} \Biggl\{  \frac12 [  \ln^2 (\frac{y_{++}}{4}) -  \ln^2 (\frac{y_{+-}}{4}) ] - 
[\ln (\frac{y_{++}}{4}) - \ln (\frac{y_{+-}}{4})] \Biggr\}
\nonumber\\
& & + \;3\; \int d^4 x' \; a^{'3} \Biggl\{  \ln (\frac{y_{++}}{4}) - \ln (\frac{y_{+-}}{4}) \Biggr\}\;,
\nonumber
\eea
With this form, the rest steps (Step 2 - Step 7) can be applied in the same way as the above example. The result is
\bea
\int d^4 x' \; a^{'3} \; f(x')
&=& 
\frac{16\pi^2 i}{H^4} \;[-\frac{\ln^2(a)}{2a}+\frac{2\ln(a)}{a}-\frac{3}{a}+\frac{\pi^2}{3a}+\mathcal{O}(\frac{\ln(a)}{a^2})]\;.
\eea 
Here the series representation of the Riemann zeta function is used.
\bea
\zeta(2) = \sum_{n=1}^{\infty} \frac1{n^2} = \frac{\pi^2}{6}\;.
\eea   


\section{Result Tables}\label{app-result-tables}

\begin{table}[htbp]
\begin{center}
\caption{$-\frac{\tilde{u}(0,k)}{a^2} \frac{iH^8}{(4\pi)^4} \int d^4x' {\rm{(Ext.\;opr)}}\times [f(\frac{y_{++}}{4}) - f(\frac{y_{+-}}{4})]$ 
from conformal nonlocal terms.}
\begin{ruledtabular}
\begin{tabular*}{1.5\textwidth}{ll}
$\phantom{sss}{\rm {External \; operators}} \times f(x)$&$\mbox{Coefficient of } \tilde{u}(0,k) \times  \frac{H^4}{(4\pi)^2}\phantom{sss}$\\[1ex]
\hline\\
$\phantom{sss}(aa')^3 \square^3/H^2 \times [\frac{9\ln{x}}{5x}] $&$0\phantom{ssssssssssssssssssssss}$\\[2ex]
$\phantom{sss}(aa')^3 \square^2 \times [\frac{9\ln{x}}{x} - \frac{21}{5x}]$&$0$\\[2ex]
$\phantom{sss}(aa')^3 H^2\square \times [-\frac{267\ln{x}}{5x} + \frac{51}{x}]$&$0$\\[2ex]
$\phantom{sss}(aa')^3 H^4 \times [\frac{258\ln{x}}{5x} - \frac{549}{5x}]$&$0$\\[2ex]
$\phantom{sss}(aa')^2(a^2+a'^2)\square^3/H^2 \times [-\frac{9\ln{x}}{10x}]$&$a \; [\frac{\ln(a)}{20}-\frac{9}{40}]$\\[2ex]
$\phantom{sss}(aa')^2(a^2+a'^2)\square^2 \times [-\frac{99\ln{x}}{10x} - \frac{9}{10x}]$&$a \;[\frac{11\ln(a)}{40}-\frac{43}{40}]$\\[2ex]
$\phantom{sss}(aa')^2(a^2+a'^2)H^2\square \times [\frac{36\ln{x}}{x} - \frac{51}{2x}]$&$a\;[-\frac{101\ln(a)}{24} + \frac{101}{48}]$\\[2ex]
$\phantom{sss}(aa')^2(a^2+a'^2)H^4 \times [-\frac{108\ln{x}}{5x} + \frac{174}{5x}]$&$a \;[\frac{3\ln(a)}{20} - \frac{83}{120}] $\\[2ex]
\hline\\
$\phantom{sss}\rm{Total}$&$a\;[-\frac{\ln(a)}{40} + \frac{9}{80}]$\\[2.5ex]
\end{tabular*}
\label{newcct4}
\end{ruledtabular}
\end{center}
\end{table}

\begin{table}[htbp]
\begin{center}
\caption{$-\frac{\tilde{u}(0,k)}{a^2} \frac{iH^8}{(4\pi)^4} \int d^4x' {\rm{(Ext.\;opr)}}\times [f(\frac{y_{++}}{4}) - f(\frac{y_{+-}}{4})]$ 
from kinetic nonlocal terms.}
\begin{ruledtabular}
\begin{tabular*}{1.5\textwidth}{ll}
$\phantom{sss}{\rm {External \; operators}} \times f(x)\;$&$ \mbox{Coefficient of } \tilde{u}(0,k) \times  \frac{H^4}{(4\pi)^2}\phantom{sss}$\\[1ex]
\hline\\
$\phantom{sss}(aa')^3 \square^3/H^2 \times [\frac{\ln{x}}{3x}]$\label{firstterm}&$0$\\[2ex]
$\phantom{sss}(aa')^3 \square^2 \times [-\frac{3\ln{x}}{2x}]$&$0$\\[2.5ex]
$\phantom{sss}(aa')^3 H^2\square \times [\frac{25\ln{x}}{x}-\frac{3}{2x}]$&$0$\\[2.5ex]
$\phantom{sss}(aa')^3 H^4 \times [-\frac{44\ln{x}}{x} - \frac{26}{x} - \ln(aa') \frac{30}{x}]$&$0$\\[2.5ex]
$\phantom{sss}(aa')^2(a^2+a'^2)\square^3/H^2 \times [-\frac{\ln{x}}{6x}]$\label{fifthterm}&$a \; [\frac{4\ln(a)}{3} - 6] $\\[2.5ex]
$\phantom{sss}(aa')^2(a^2+a'^2)\square^2 \times [\frac{7\ln{x}}{6x} - \frac{1}{6x}]$&$a \; [-\frac{14\ln(a)}{3} + \frac{58}{3}]$\\[2.5ex]
$\phantom{sss}(aa')^2(a^2+a'^2)H^2\square \times [\frac{\ln{x}}{3x} + \frac{5}{6x}]$&$ a\; [-\frac{4\ln(a)}{3} + \frac{26}{3}]$\\[2.5ex]
$\phantom{sss}(aa')^2(a^2+a'^2)H^4 \times [-\frac{4\ln{x}}{x} - \frac{7}{2x}]$&$a \; [ 4 \ln(a) -\frac{34}{4}] $\\[2.5ex]
\hline\\
$\phantom{sss}\rm{Total}$&$ a\; [-\frac{2\ln(a)}{3} + \frac{51}{2} ] $\\[2.5ex]
\end{tabular*}
\label{newkt4}
\end{ruledtabular}
\end{center}
\end{table}


\begin{table}[htbp]
\begin{center}
\caption{$-\frac{\tilde{u}(0,k)}{a^2} \frac{iH^8}{(4\pi)^4} \int d^4x' {\rm{(Ext.\;opr)}}\times [f(\frac{y_{++}}{4}) - f(\frac{y_{+-}}{4})]$ 
from cross nonlocal terms.}
\begin{ruledtabular}
\begin{tabular*}{1.5\textwidth}{ll}
$\phantom{sss}{\rm {External \; operators}} \times f(x)$&$\mbox{Coefficient of } \tilde{u}(0,k) \times  \frac{H^4}{(4\pi)^2}\phantom{sss}$\\[1ex]
\hline\\
$\phantom{sss}(aa')^3 \square^3/H^2 \times [\frac{2\ln{x}}{x}]$&$0$\\[2ex]
$\phantom{sss}(aa')^3 \square^2 \times [-\frac{27\ln{x}}{x} ]$&$0$\\[2.5ex]
$\phantom{sss}(aa')^3 H^2\square \times [ -\frac{378\ln{x}}{x} - \frac{59}{x}]$&$0$\\[2.5ex]
$\phantom{sss}(aa')^3 H^4 \times [\frac{608\ln{x}}{x} - \frac{436}{x}]$&$ 0$\\[2.5ex]
$\phantom{sss}(aa')^2(a^2+a'^2)\square^3/H^2 \times [\frac{\ln{x}}{x}]$&$  a \; [-8 \ln(a) +36]$\\[2.5ex]
$\phantom{sss}(aa')^2(a^2+a'^2)\square^2 \times [-\frac{28\ln{x}}{x} + \frac{1}{x}]$&$ a\; [112 \ln(a) -452]$\\[2.5ex]
$\phantom{sss}(aa')^2(a^2+a'^2)H^2\square \times [\frac{178\ln{x}}{x} - \frac{26}{x}]$&$ a\; [-712\ln(a) +2952]$\\[2.5ex]
$\phantom{sss}(aa')^2(a^2+a'^2)H^4 \times [-\frac{252\ln{x}}{x} - \frac{66}{x}]$&$ a \;[128\ln(a) -702]$\\[2.5ex]
\hline\\
$\phantom{sss}\rm{Total}$&$a\;[-352 \ln(a) + 1834] $\\[2.5ex]
\end{tabular*}
\label{newct4}
\end{ruledtabular}
\end{center}
\end{table}
In the Tables \ref{newcct4}, \ref{newkt4} and \ref{newct4}, 
the zeros are not exactly zero but sub-dominant to the leading terms. 
Below we give examples in Appendix \ref{app-example-compute-result} for how the result Tables were made.

\section{The Examples From The Table~ \ref{newkt4}}\label{app-example-compute-result}

\begin{itemize}
\item{
We set zero for the following sub-dominant terms:

For example, the first term in Table \ref{newkt4} is
\bea
(aa')^3 \frac{\square^3}{H^2} \times [\frac{\ln(x)}{3x}] &=& a^3 \square^3 \int d^4 x' a'^3 \;[\frac{\ln(x)}{3x}]\;,
\nn\\
&=& \frac13 a^3 H^4 \square^2 \;[- \frac{\ln(a)^2}{2a} + \frac{2\ln(a)}{a}  -\frac{3}{a} + \frac{\pi^2}{3a}  ]\;,
\nn\\
&=& a^2 H^4  [ -\frac23 \ln(a)^2 + 4\ln(a) - \frac{17}{3} + \frac49 \pi^2 ]\;,
\nn\\
\frac{1}{a^2}(aa')^3 \frac{\square^3}{H^2} \times [\frac{\ln(x)}{3x}] &=& H^4  [ -\frac23 \ln(a)^2 + 4\ln(a) - \frac{17}{3} + \frac49 \pi^2 ] \rightarrow 0\;,
\eea 
That is, we set zero for the terms of order $\mathcal{O}(\ln(a)^2)$ because they are sub-dominant to $a\ln(a)$ and $a$.
}

\item{
Leading terms we keep:

For example, the fifth term in Table \ref{newkt4} is
\bea
(aa')^2 (a^2 + a'^2) \frac{\square^3}{H^2} \times [-\frac{\ln(x)}{6x}] &=& 
-\frac16 a^2 \frac{\square^3}{H^2} \int d^4 x' a'^4 \frac{\ln(x)}{x} - \frac16 a^4 \frac{\square^3}{H^2}  \int d^4 x' a'^2 \frac{\ln(x)}{x}\;,
\nn\\
&=& -\frac16 a^2 \frac{\square^3}{H^2} (\frac34) - \frac16 a^4 \frac{\square^3}{H^2} [-\frac{\ln(a)}{a} + \frac3{a}]\;,
\nn\\
\frac{1}{a^2}(aa')^2 (a^2 + a'^2) \frac{\square^3}{H^2} \times [-\frac{\ln(x)}{6x}] &=& 
H^4 a\;[\frac43\ln(a) - 6 ]\;.
\eea
}
\end{itemize}


\end{document}